# Single Spin Logic Implementation of VLSI Adders

Soumitra Shukla and Bahniman Ghosh
Department of Electrical Engineering, Indian Institute of Technology, Kanpur, India
Email: smtrshuk@iitk.ac.in

**Abstract:** Some important VLSI adder circuits are implemented using quantum dots (qd) and Spin Polarized Scanning Tunneling Microscopy (SPSTM) in Single Spin Logic (SSL) paradigm. A simple comparison between these adder circuits shows that the mirror adder implementation in SSL does not carry any advantage over CMOS adder in terms of complexity and number of qds, opposite to the trend observed in their charge-based counterparts. On the contrary, the transmission gate adder, Static and Dynamic Manchester carry gate adders in SSL reduce the complexity and number of qds, in harmony with the trend shown in transistor adders.

## I. INTRODUCTION

The VLSI technology in the present scenario is confronted by the famous Red Brick Wall problem [1]. This problem has set a deadlock to the trend of miniaturization of devices in this field. As a result, the researchers have started to look for alternative devices, circuits and methods to overcome this deadlock. One of the approaches becoming popular these days uses quantum dots, electrons and magnetic field. Together this is called Single Spin Logic paradigm (SSL) [2-6]. The reason for SSL becoming popular is because the spin state of the electron remains stable [7] in qds as compared to bulk devices and as a result implementing them becomes easier [8-11] .With the exploration and exploitation of this idea in the recent times[12,13], researchers have come up with techniques that demonstrate how to implement some fundamental digital circuits like NAND gate [1,8], MUX [7],AND OR EXOR[9], ALU [2] etc. Using the same prototype we are trying to extend this idea and applying it to some more fundamental circuits, particularly Full Adders.

The circuits belonging to this class show the same behavior in terms energy dissipation as in the normal VLSI based circuits i.e. the quality metric of Energy-Delay product shows the tradeoff between energy supplied and speed of response unlike what is expected in quantum computers and quantum gates [14]. A quantum computer is a new device idea that uses principles of quantum mechanics for data processing. The basic components of a quantum computer are quantum gates that are supposed to be non-dissipative in terms of energy. Further the implementation of NAND gate has been shown in FIG.1 using Quantum dots by arranging them in a particular fashion [1,8]. The single electron in each qd interacts with the other via nearest–neighbor exchange interaction [2,3,5] .The origin of this interaction is purely quantum mechanical in nature. By stating the nearest-neighbor interaction we make sure that electrons only in the adjacent cells (qd) interact. A consequence of this assumption is that the logic signal can travel to different nodes through these interactions [14] with the help of supporting clock pads. These clock pads are driven by a 3-phase clock.

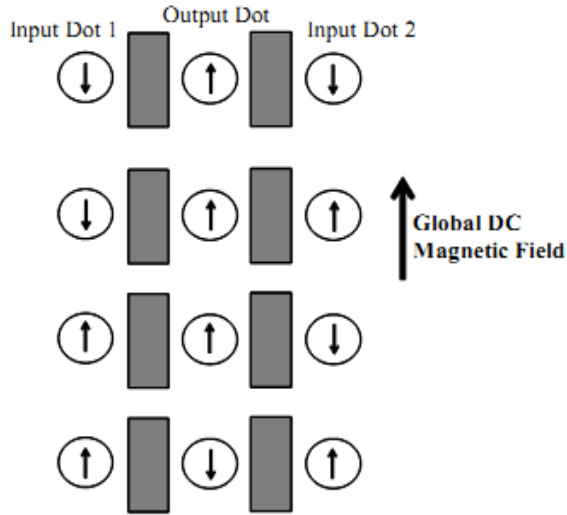

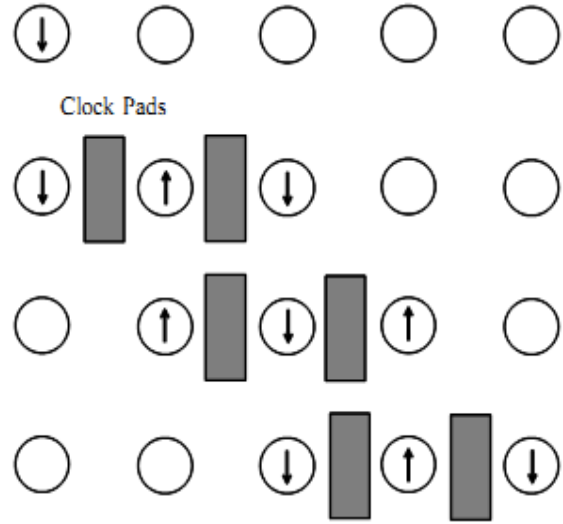

FIG.1. Example of a NAND gate implementation in the SSL paradigm. The rectangles in between the dots are the clock pads

This clock scheme serves two purposes; one it ensures that communication of signal is from input to output and not in reverse direction [2], and the other it makes the structure pipelined, endorsing high speed. However, the clock pads are not shown in the figure for the sake of clarity. In general for building a circuit in SSL paradigm proper planning for the layout of the QD's in a 2D array and correct pin configuration for inputs and outputs is required. Further, the Truth table of a Full Adder is shown in Table 1.with A, B, Ci as inputs and SUM (S) and Carry (Co) as outputs.

**Table 1.**

| `A | B | Ci | S | Co |
|---|---|---|---|---|
| 0 | 0 | 0 | 0 | 0 |
| 0 | 0 | 1 | 1 | 0 |
| 0 | 1 | 0 | 1 | 0 |
| 0 | 1 | 1 | 0 | 1 |
| 1 | 0 | 0 | 1 | 0 |
| 1 | 0 | 1 | 0 | 1 |
| 1 | 1 | 0 | 0 | 1 |
| 1 | 1 | 1 | 1 | 1 |

FIG.2. Figure showing the use of clock scheme for transferring logic through quantum dots

## II. Complementary CMOS Adder

Simple VLSI circuit of complementary CMOS Adder is shown in FIG. 3. Same circuit is implemented in SSL and shown in FIG.4. As stated earlier one of the requirements for SSL is magnetic field and carefully patterned two dimensional arrays of electron cells. The global Magnetic Field is pointing in the upward direction, and with respect to this Magnetic Field the 'up-spin' of electrons in cells represents logic HIGH (i.e. 1), and similarly 'down-spin' represents logic LOW (i.e. 0). The big block of pointed of arrow in the figure is the SPSTM (Spin Polarized Scanning Tunneling Microscope) tip [6, 15], which is used to 'read' and 'write' the spin in the associated electron cell. However, other techniques have also been demonstrated for reading the electron spin [16-18]. The electrons in the two adjacent electron cell couple via nearest-neighbor exchange interaction. The interaction of two isolated electron cells makes the electrons in the two cells to be in opposite state of spin polarization (i.e.'1, 0' or '0, 1').When a single cell comes under the interaction of two different electron cell having different independent parent SPSTM tip, then the former cell generates the NAND logic output. This principle is used to build these 2D-arrays of electron cells

In FIG. 5 a simple case of the Truth Table of the Single-Bit Full Adder is shown. Take for example Column 5 of Table 1, i.e. A=1(up-spin), B=0(down-spin), Ci=1(up-spin); in this case we read the output at the cell with which SPSTM with name 'S' is attached. We get SUM, S=0(down-spin) and Carry Output, Co=1(up-spin). These values of output are also verified by the Table 1. Using the same argument the other versions of adders are realized

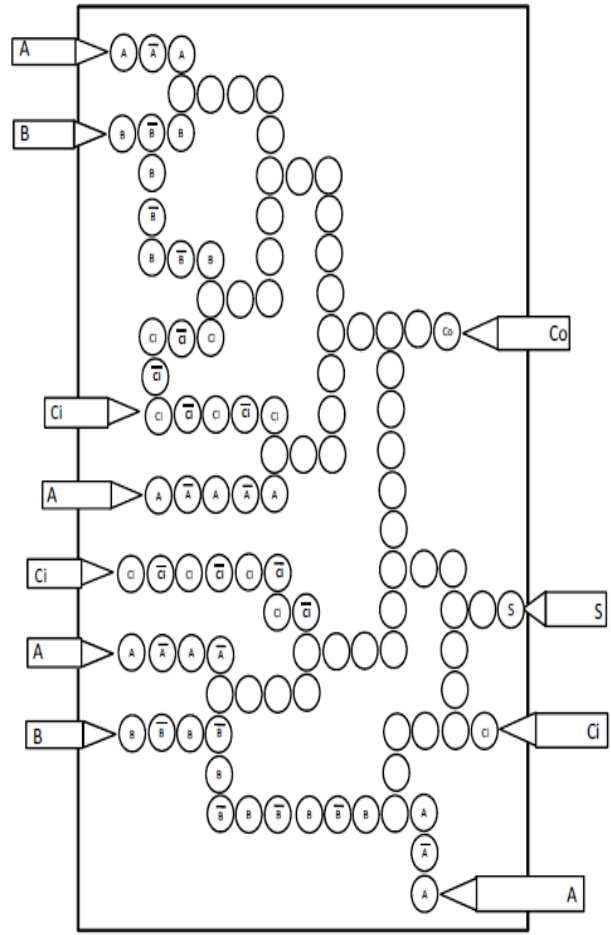

FIG.4. A Full adder circuit designed in SSL.

### III. Mirror Full Adder

The Mirror Adder circuit, shown in FIG.6 is a very important circuit because of its numerous advantages over complementary CMOS adder. One of the advantages is its symmetrical circuit topology. Transistor based Mirror Adder is different because of its functionality i.e. the pull-up and pull-down network path through which value of the logic is transferred to the outputs $\overline{SUM}$ and $\overline{CARRY}$ are different from CMOS adder. However, each path is still mutually exclusive i.e. only one path active at a time. Using the same principle to build the Full Adder with electron cells we need a means to couple and decouple the interaction between any

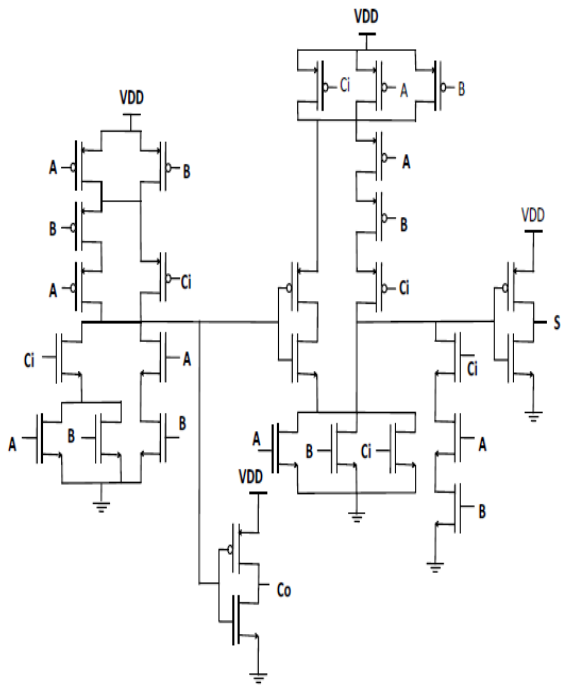

FIG 3 Complementary CMOS Full Adder

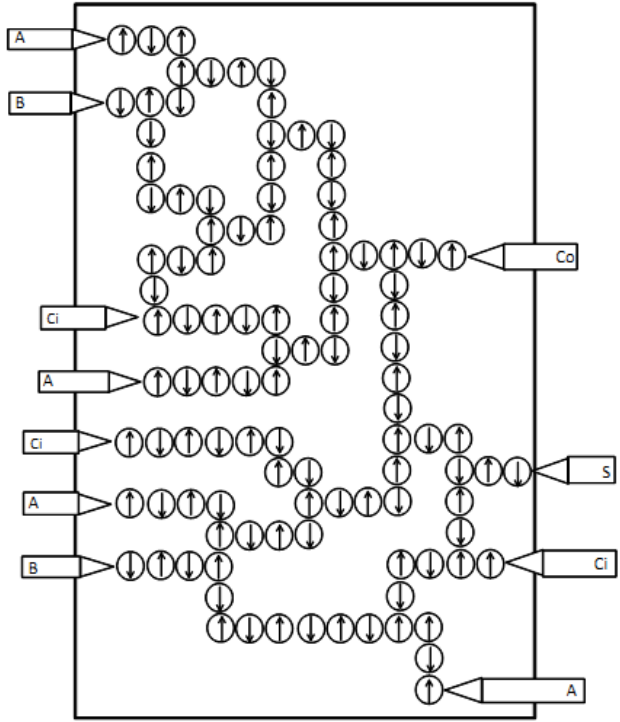

FIG.5. Example showing function of the circuit as a full adder i.e. for inputs A=1, B=0, Ci=1, the value at the outputs are S=0, Co=1.

Two particular cells depending on the input pattern [4]. For this purpose the gates are used in between cells. However, the difference between these gates and clock pads is that clock pads are controlled by external means but the gate pads should be internally controlled by a qd. In this case as shown in the Fig. 7 gates at cell Co is controlled by (A.B), ($\overline{A}.\overline{B}$), and SUM(S) is controlled by gate-3(A ⊕ B ⊕ Ci), gate-1(A.B.C), gate-2 ($\overline{A}.\overline{B}.\overline{C}$). When observed carefully gate-3 is controlled by a qd which contains the value of logic (A ⊕ B ⊕ Ci), which in turn is the value of the SUM output. This indicates that implementing Mirror Adder may not be wise in SSL.

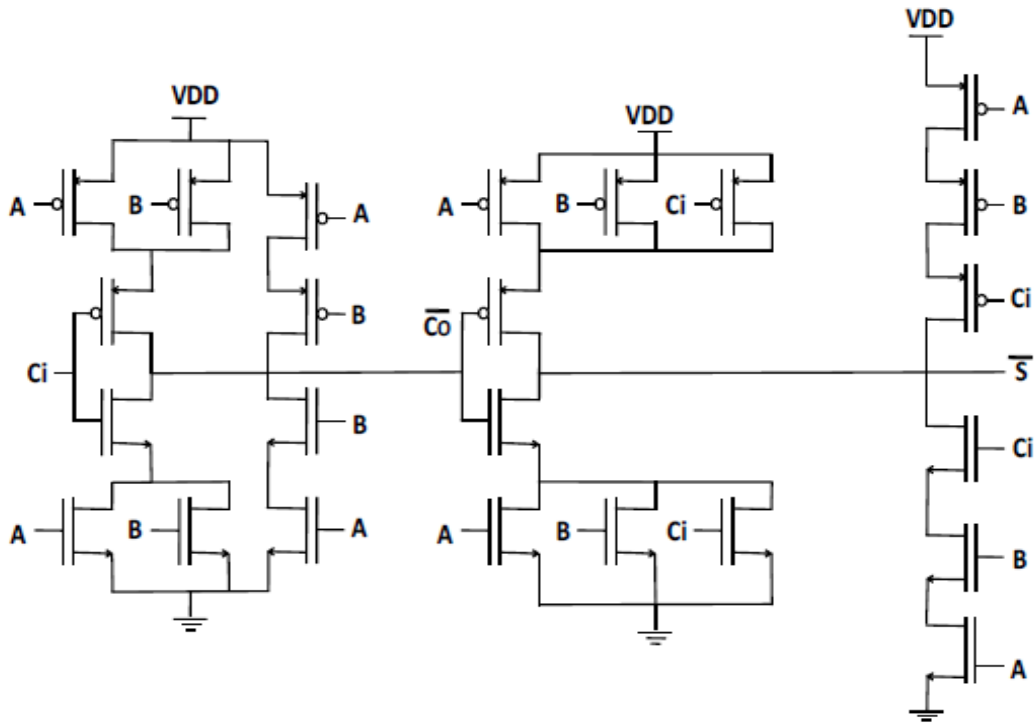

FIG 6. Mirror adder design, full adder circuit

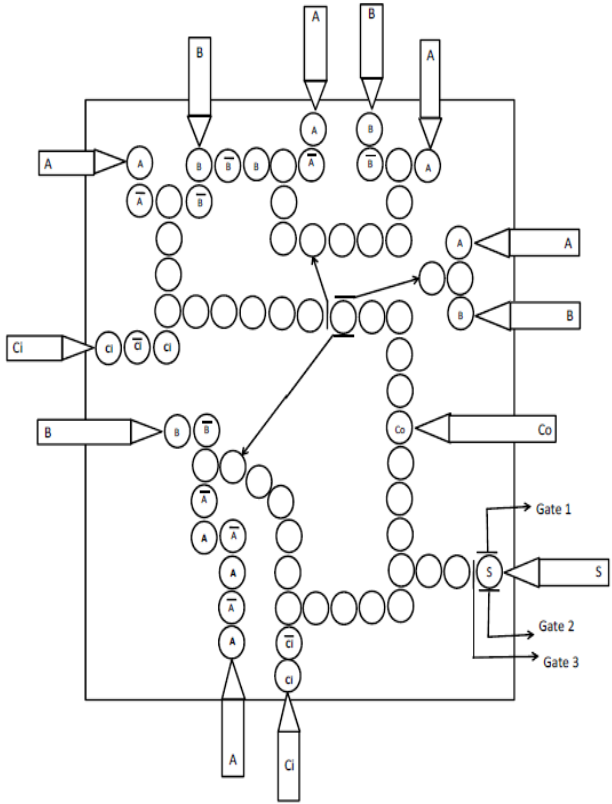
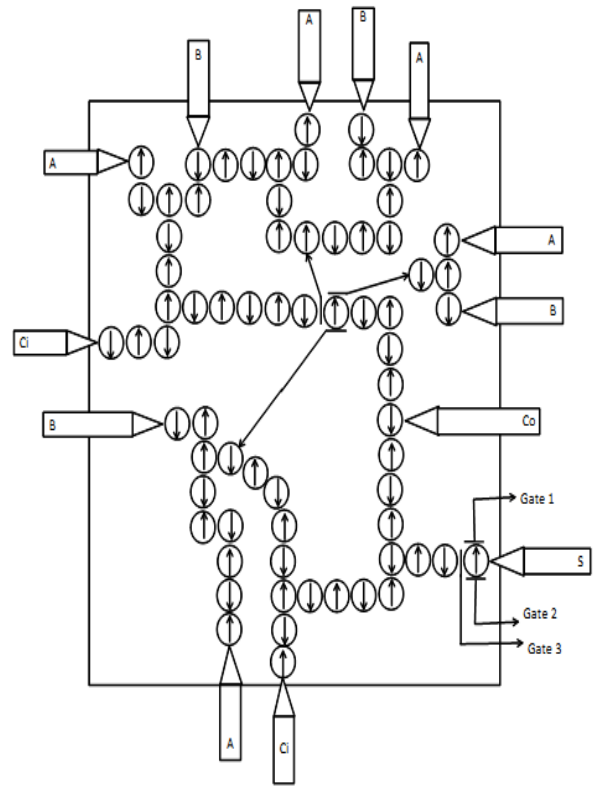

FIG. 7. Mirror Adder circuit implementation in SSL

FIG. 8. Example showing the Mirror Adder circuit, as an adder with inputs A=1, B=0, Ci=0 and outputs S=1, Co=0

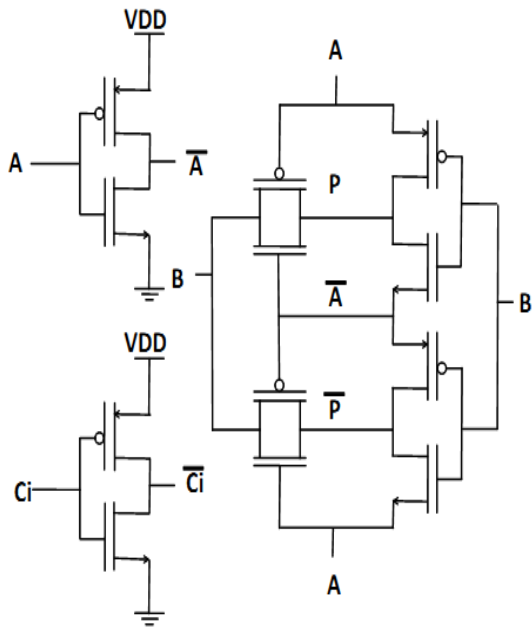
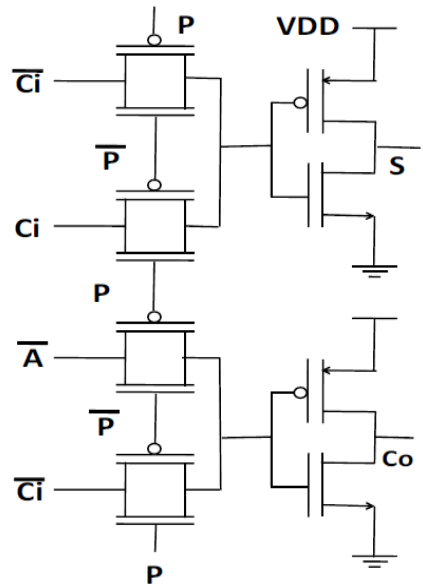

Fig 9.   Setup circuit for transmission gate

Fig 10. Transmission gate adder

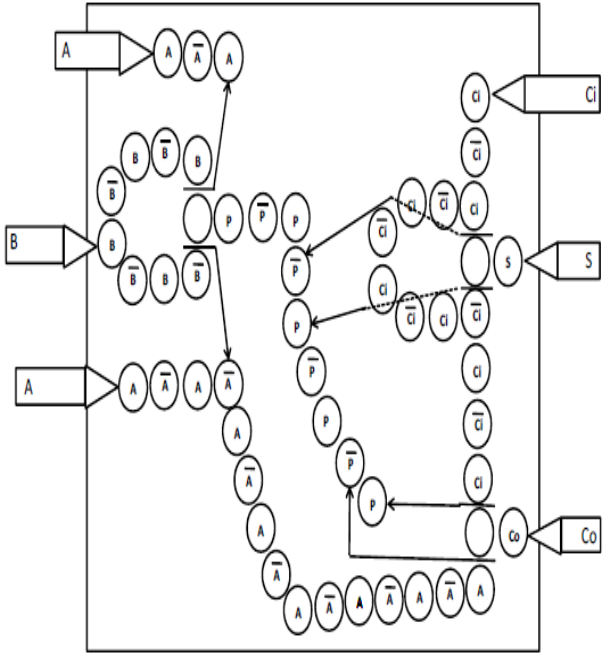

FIG 11 Transmission gate adders in SSL.

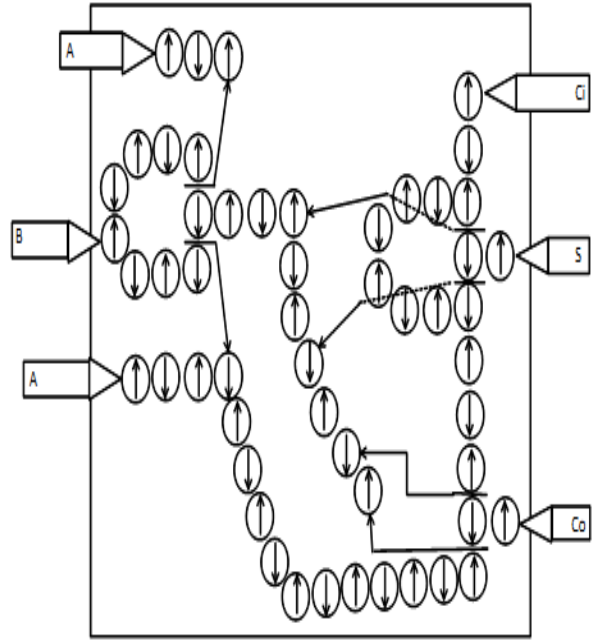

FIG 12. Example showing the Transmission gate as an adder, with inputs A=1, B=0, Ci=0 and outputs S=1, Co=0.

### IV. Transmission Gate Adder

The circuit for Transmission gate as shown above in FIG.9, requires a 'setup' sub-circuit to produce intermediate signals like propagate signal (P), which is to be used subsequently in the final circuit for getting the SUM and CARRY outputs. In Fig.9, When A=0, P=B and when A=1, P=$\bar{B}$ similarly for the final circuit, when P=0, S=Ci, and when P=1, S=$\bar{Ci}$. Likewise, for the carry output when P=0, Co=A and when P=1, Co=$\bar{Ci}$. . Same principle is utilized for getting the outputs while implementing the circuit in SSL paradigm in FIG 11.

### V. Static Manchester Carry Chain Gate

The technique of getting SUM output in a Static Manchester Carry gate is similar to the Transmission gate adder, the only difference is in generation of CARRY output. In this scheme there are three different signals Propagate ( A $\oplus$ B), Generate (A. B) and Delete ($\bar{A}$.$\bar{B}$). Here the value assigned to CARRY comes from three different mutually exclusive paths, depending on the switching operation performed by the control signals which are applied at the gate terminal of the transistors as shown in FIG 13.

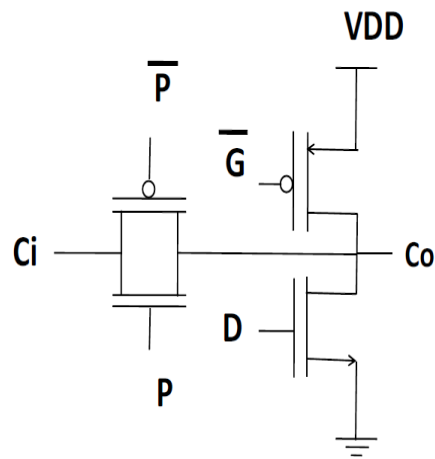

FIG.13. Carry generation using the principle of static Manchester Carry gate.

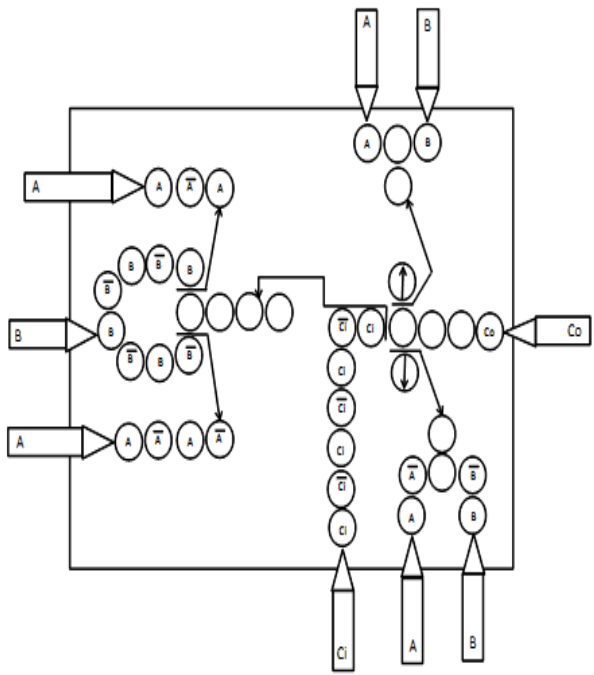

FIG 14. Static Manchester Carry gate implemented using SSL.

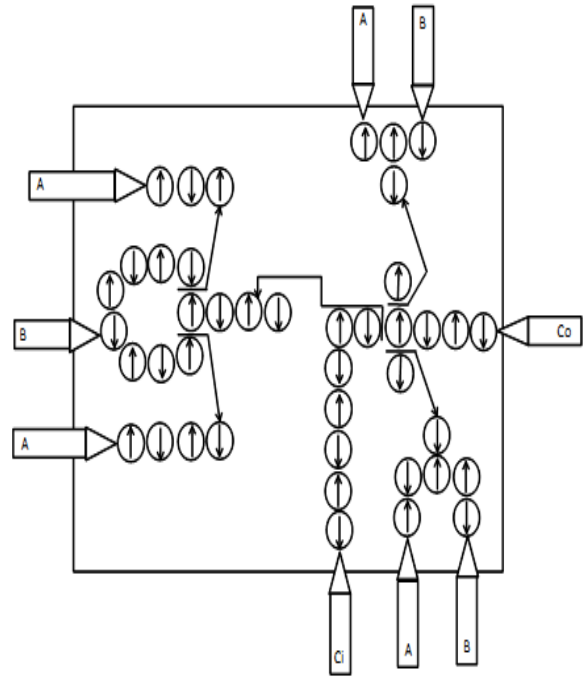

FIG 15. Example showing the Static Manchester Carry circuit as an adder with A=1,B=0,Ci=0 and outputs Co=0.

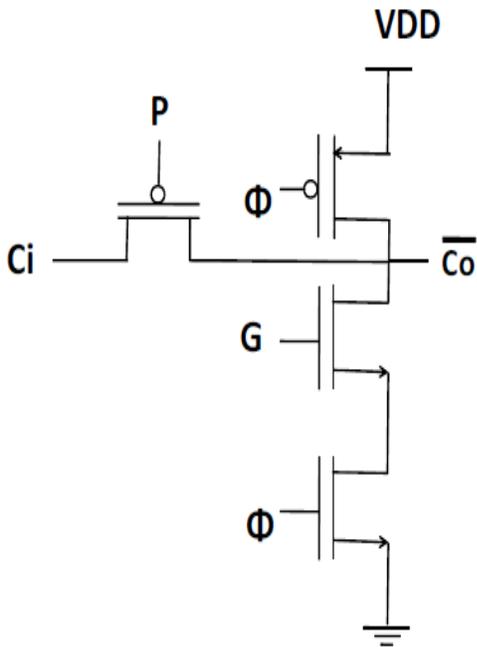

FIG.16. Dynamic Manchester Carry gate

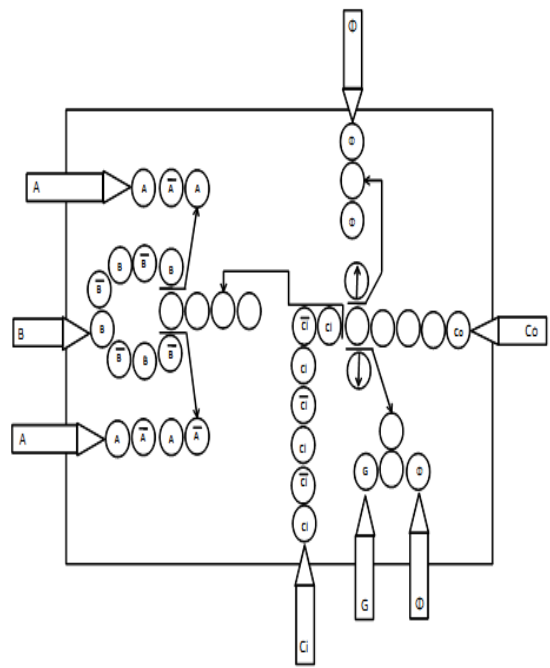

FIG 17. Dynamic Manchester Carry gate implemented using SSL.

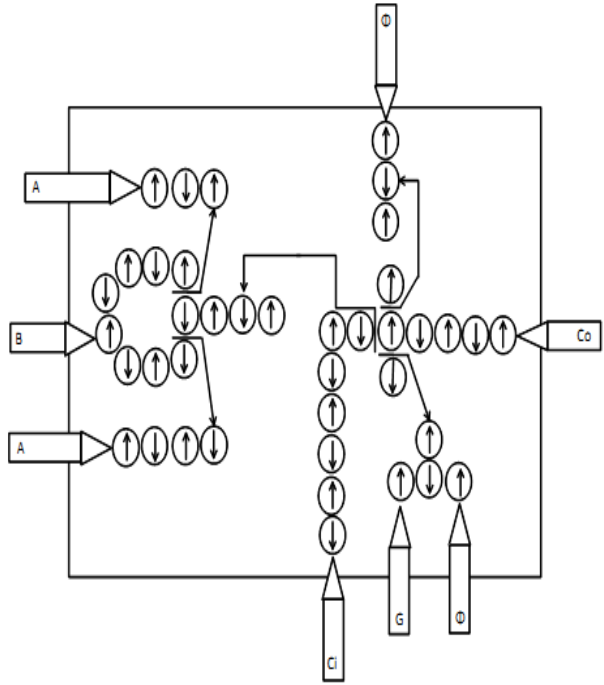

FIG 18. Example showing the Dynamic Manchester Carry gate as an adder with clock high and A=1,B=1,Ci=0 and outputs Co=1.

## VI.  Dynamic Manchester Carry Gate

The dynamic Manchester carry gate is slightly different from the static Manchester carry gate in terms of the circuit topology. The reason for this difference is because of its dynamic nature i.e., the presence of clock signal. However, the basic operation is similar to static Manchester carry gate. The signals used are also same except Delete ($\overline{A}.\overline{B}$) is replaced by the clock ($\Phi$). Further, as in the case of Transmission gate adder, only the carry generation circuit is shown because SUM generation circuit is same.

## VII.  RESULTS AND DISCUSSION

We have tried to implement the full adder circuits in the paradigm of SSL, maintaining the functional behavior of the circuit as is preserved in its transistor implementation. Further, the implementation of a Complementary CMOS Adder is very straight-forward. However, the designing of other adders became slightly complicated and enabled us to use the idea of gate pads [*Note: The term gate pads is used to differentiate them from the clock pads*] to preserve the functional behavior of the circuits. As a result Mirror adder, Transmission gate adder and Manchester Carry Chain adders are implemented in the same paradigm by using gate pads controlled by qds.

## VIII.  CONCLUSION

We conclude that the implementation of Mirror adder circuit introduces redundancy and complexity by introducing extra gates within the structure in the SSL implementation, without any change in the count of quantum dots. This observation is just opposite of what is seen in charge-based circuits where the former has many advantages over the latter. Further, the transmission gate and both the manchester carry chain adders reduce the number of quantum dots to reasonable extent same as in their charge-based counterparts.

The reason behind this observation is that the transistor-based circuits shown can be primarily classified into two types, depending on their functional behavior, one in which the transistor network present in the circuit work together to generate the desired logic, and second in which transistors work as a switch, making the nature of the circuit a DATAFLOW type. These DATAFLOW type of circuits when implemented in the single-spin paradigm show a reduction in number of Quantum Dots, when implemented using the gate pads

**Referances**

[1] Rabaey, Jan M., Nikolic, Borivoje,:"Digital Integrated Circuits A Design Perspective", Second Edition

[2] Bandopadhyay, Supriyo,Cahay ,Marc,:'TOPICAL REVIEW Electron spin for classical information processing: a brief survey of spin-based logic devices, gates and circuits', *Nanotechnology* 20(2009) 412001(35pp)

[3] Basu , T.,Sarkar,S.K.,Bandopadhyay, S.,:'Arithmetic logic unit of a computer based on the spin-polarised single electrons', QUANTUM DOTS AND NANOWIRES, *IET Circuits Devices Syst.*, Vol . 1, No. 3, June 2007

[4] Bhattacharya, T.K., Neogy, A.,Bhowmik, D.:'Single-Spin Implementation of Multiplexer', *Physica* E 41 (2009) 1184-1186

[5] Bandopadhyay, Supriyo., Cahay ,Marc,: 'Introduction to spintronics'

[6] Bandopadhyay, S.,Das, B.,and Miller, A.E:'Supercomputing with spin polarized single electrons in a Quantum coupled architecture', *Nantechnology.*, 1994, 5,pp. 113-133

[7] A.V. Khaetskii and Y.V. Nazarov, Phys. Rev. B 61, 12 639 (2000)

[8] Privman, V., Mozyrsky, D., and Wagner, I.D.: 'Quantum computing with spin qubits in semiconductor structures', *Comput. Phys.Commun.*, 2002, 146, pp. 331–338

[9] Kane, B.E.: 'A silicon based nuclear spin quantum computer', *Nature*,1998, 393, pp. 133–137

[10] Loss, D., and DiVincenzo, D.P.: 'Quantum computation with quantum dots', Phys. Rev. A, 1998, 57, pp. 120 – 126

[11] Bandyopadhyay, S.: 'Self assembled nanoelectronic quantum computer based on the Rashba effect in quantum dots', Phys. Rev.B, 2000, 61, pp. 13813 – 13820

[12] Craig, N.J., Taylor, J.M., Lester, E.A., Marcus, C.M., Hanson, M.P.,and Gossard, A.C.: 'Tunable non-local spin control in a coupled quantum dot system', *Science*, 2004, 304, pp. 565–567

[13] Livermore, C., Crouch, C.H., Westervelt, R.M., Campman, K.L., and Gossard, A.C.: 'The Coulomb blockade in coupled quantum dots',*Science*, 1996, 274, pp. 1332–1335

[14] Bychkov, A.M., Openov, L.A. and Semenihin, I.A.: 'Single electron computing without dissipation', JETP Lett., 1997, 66,pp 298-303

[15] Molotkov, S.N., and Nazin, S.N.: 'Single electron spin logical gates', *JETP Lett.*, 1995, 62, pp. 273–281

[16] Rugar, D., Budakian, R., Mamin, H.J., and Chui, B.W.: 'Single spin detection by magnetic resonance force microscopy' , *Nature* ,2004,430, pp.329-332

[17] Xiao, M., Martin, I., Yablonovitch, E., and Jiang, H.W.: 'Electrical detection of the spin resonance of a single electron in a silicon field effect transistor', Nature, 2004, 430, pp. 435 – 439

[18] Elzerman, J.M., Hanson, R., Beveren, van,Witkamp, B., Vanderyspen, L.M.K., and Kouwenhoven, L.P.:'Single shot read out of an individual electron spin in a quantum dot', *Nature*, 2004, 430,pp 431-435.

.